\documentclass[pdflatex,sn-standardnature]{sn-jnl}
\theoremstyle{thmstyleone}%
\theoremstyle{thmstyletwo}%

\theoremstyle{thmstylethree}%

\usepackage{afterpage}
\usepackage{ulem}
\usepackage{bm}
\usepackage{lineno}
\hypersetup{bookmarksdepth=3} 

\usepackage{caption}
\DeclareCaptionLabelFormat{extended}{\textbf{Extended Data Fig. #2}}

\begin{document}

\title[A model for heating the super-hot corona in solar active regions]{A model for heating the super-hot corona in solar active regions}


\author[1,2]{\fnm{Zekun} \sur{Lu}}

\author*[1,2]{\fnm{Feng} \sur{Chen}}\email{chenfeng@nju.edu.cn}

\author*[1,2]{\fnm{M. D.} \sur{Ding}}\email{dmd@nju.edu.cn}

\author[1,2]{\fnm{Can} \sur{Wang}}

\author[1,2]{\fnm{Yu} \sur{Dai}}

\author[1,2]{\fnm{Xin} \sur{Cheng}}

\affil[1]{\orgdiv{School of Astronomy and Space Science}, \orgname{Nanjing University}, \orgaddress{\street{163 Xianlin Road}, \city{Nanjing}, \postcode{210023}, \state{Jiangsu}, \country{China}}}

\affil[2]{\orgdiv{Key Laboratory of Modern Astronomy and Astrophysics (Nanjing University)}, \orgname{Ministry of Education}, \orgaddress{\street{163 Xianlin Road}, \city{Nanjing}, \postcode{210023}, \state{Jiangsu}, \country{China}}}


\abstract{What physical mechanisms heat the outer solar or stellar atmosphere to million-Kelvin temperatures is a fundamental but long-standing open question. In particular, the solar corona in active region cores contains an even hotter component reaching ten million Kelvin, manifesting as persistent coronal loops in extreme ultraviolet and soft X-ray images, which imposes a more stringent energy budget. Here, we present a self-consistent coronal heating model using a state-of-the-art three-dimensional radiative magnetohydrodynamics simulation. We find that the continuous magnetic flux emergence in active regions keeps driving magnetic reconnections that release energy impulsively but, on time average, persistently. As a result, numerous sub-structures are heated to ten million Kelvin and then evolve independently, which collectively form long-lived and stable coronal loops as in observations. This provides a heating model explaining the origin of the super-hot coronal plasma and the persistence of hot coronal loops in emerging active regions. 
}

\maketitle

\section{Introduction}\label{sec:intro}

The solar corona, the outer atmosphere of the Sun, is found at million-Kelvin temperatures, more than one hundred times hotter than the underlying photosphere. Under such temperatures, coronal plasma is suffering from an energy loss through heat conduction and radiation, which requires one or more heating mechanisms to constantly heat the corona to maintain the energy balance. However, although continuously debated, no consensus has been reached yet on this long-lasting {\it{coronal heating problem}} \cite{Klimchuk-2006SoPh..234...41K,Parnell-De-Moortel-2012RSPTA-Review.370.3217P}. What further complicates the problem is that the coronal plasma is distributed inhomogeneously, of which the hottest components usually appear in cores of active regions. Through integrating the observations in wavelengths of ultraviolet (UV; \cite{Parenti-2017ApJ-SUMMER-UV...846...25P}), extreme ultraviolet (EUV; \cite{Warren-2012ApJ-survey...759..141W,Brosius-2014ApJ-FeXIX...790..112B}), soft X-Ray (SXR; \cite{Reale-2009ApJ...698..756R,Sylwester-2010A&A-RESIK-SXR...514A..82S,Caspi-2015ApJ-SXR-Amptek-X123-SDD...802L...2C}) and hard X-ray (HXR; \cite{Schmelz-2009bApJ...704..863S,Ishikawa-2017NatAs-FOXSI-2...1..771I}), extensive differential emission measure (DEM) analyses show that, even at a long timescale of hours \cite{Porter-1995ApJ...454..499P,Warren-2011ApJ-six-hours...734...90W}, such hotter plasmas are in high temperatures peaking around 3-4 MK, with a non-negligible super-hot portion around 10 MK. Since the conductive loss is almost proportional to the cubic of temperature, we are facing a more challenging problem: what physical processes continuously heat these super-hot coronal plasma? 

Thanks to the advances of observational capacities, considerable new insights into the coronal heating mechanisms have been provided. For example, magnetohydrodynamic (MHD) waves \cite{Van-2020SSRv-MHD-heating..216..140V} and spicules \cite{De-Pontieu-2011Sci-spicule-heating...331...55D,Samanta-Tian-2019Sci-spicule...366..890S,Bose-2023ApJ-CBP-spicule...944..171B} have been found to be ubiquitous in the solar atmosphere and supplying energy for heating the quiet corona. Nevertheless, it requires further investigations whether they can adequately heat the hotter corona in active regions \cite{Tian-2021SoPh-upflow..296...47T,Yuan-2023NatAs...7..856Y}.

An alternative scenario is that the corona is heated by a number of energy release events caused by magnetic reconnections. In this proposal, each heating event is intense and short-lived, leading to the formation of an isothermal fine structure known as a coronal strand; the observed coronal loop consists of numerous such strands. The thermodynamic models of this mechanism can reproduce reasonable thermal and radiative properties \cite{YangKai-2018NatCo...9..692Y}, especially existence of the super-hot component \cite{Cargill-2015RSPTA-nanoflare.37340260C,Barnes-2016ApJ-nanoflare-single...829...31B}, of the corona in cores of active regions. However, these studies are mostly based on an assumption of {\it{ad hoc}} heating rate, leaving uncertainties of how the magnetic energy is released. For instance, besides the hypothesis that the energy is released in nanoflares produced by magnetic braiding \cite{Parker-1983ApJ...264..642P,Parker-1988ApJ...330..474P,Cirtain-2013Natur.493..501C,Chitta-2022A&A...667A.166C}, recent studies find that magnetic cancellations near the footpoints of magnetic fields with different polarities can also power the solar corona \cite{Chitta-2017ApJS..229....4C,Priest-Chitta-2018ApJ...862L..24P,Cooper-2021MNRAS-NuSTAR.507.3936C}. Thus, a self-consistent picture of how the magnetic field evolves, heats the coronal plasma and maintains its long-term observational appearance is still lacking.

Here, we present a comprehensive three-dimensional (3D) radiative MHD simulation that has several merits. First, the simulation covers a large height range from the convective zone to the corona. By coupling with a global-scale dynamo simulation at the bottom boundary, the simulation is able to self-consistently depict the generation, transport and dissipation of magnetic energy through different layers of the solar atmosphere. Second, the physical processes that are vital to the energy transport and deposition in the solar atmosphere, e.g. radiative transfer in the photosphere and optically thin radiative cooling in the transition region and the corona, as well as field-aligned thermal conduction, are taken into account, so that the simulation reproduces more realistic thermodynamic properties of the plasma and yields observables of active regions matching those in remote sensing observations \cite{Rempel-2017ApJ...834...10R,Cheung-2019NatAs...3..160C}.

\section{Hot coronal loops in the simulated active region}\label{sec:observation}

\begin{figure}[ht]
    \centering
     \includegraphics[width=1.0\textwidth]{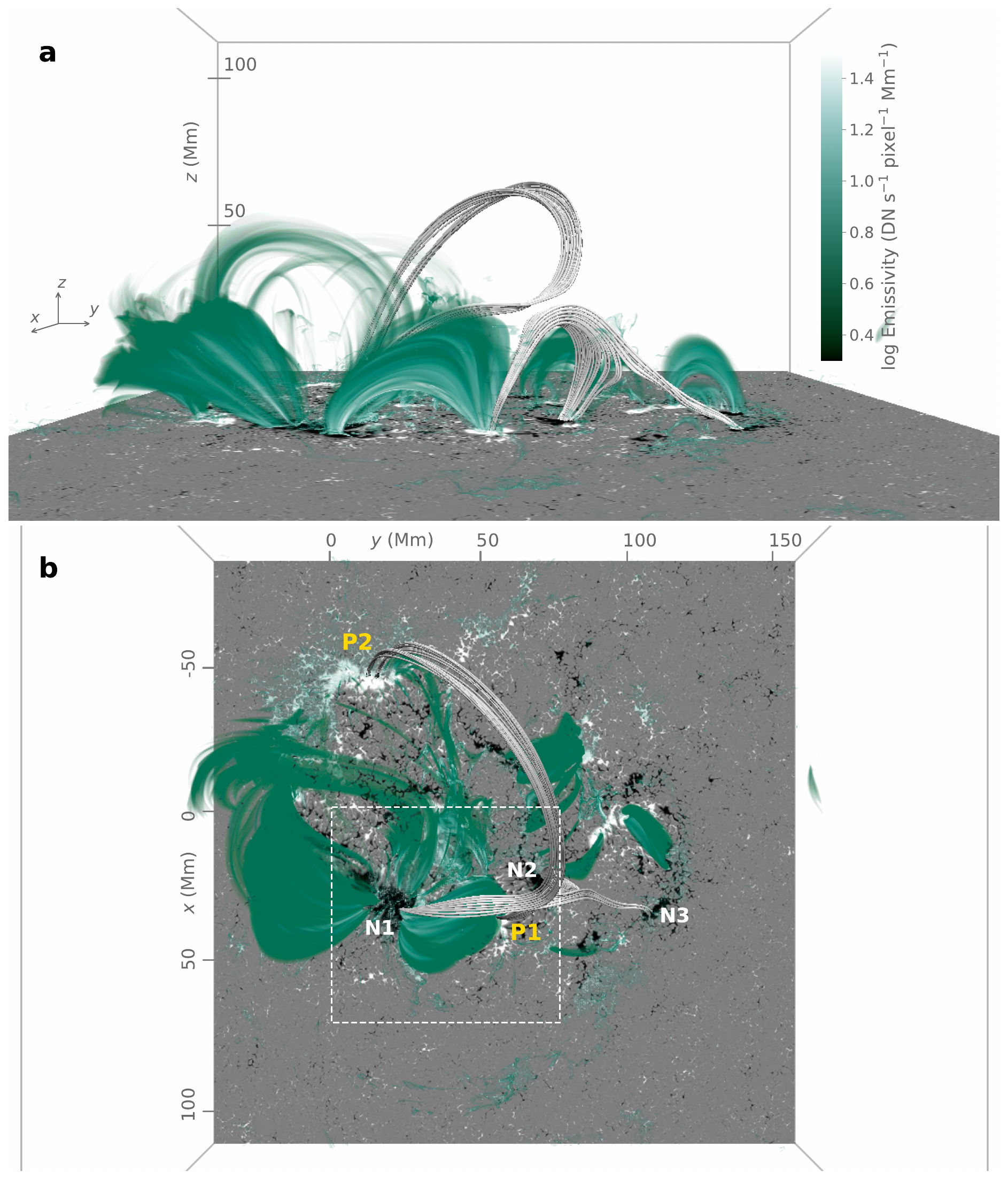}
    \caption{\textbf{The 3D modelling of EUV coronal loops in active region cores.} \textbf{a}, An $x$-direction view of the synthetic emission of the 3D modelling in AIA 94 \AA, an EUV band sampling plasma at temperatures of $\sim$ 6.3 MK, at $t = 7.2$ min, where the domain spans $x=[-86.4,110.208]$ Mm, $y=[-38.4,158.208]$ Mm, and $z=[-9.6,113.28]\ \mathrm{Mm}$. The white solid tubes represent magnetic field lines and the bottom surface is the magnetogram ($B_z$) of the photosphere. \textbf{b}, A $z$-direction view of the synthetic emission in AIA 94 \AA. P1 and P2 denote the positive polarities involved in magnetic reconnections, while N1, N2, and N3 denote the corresponding negative polarities. The white dashed box indicates the region shown in Fig. \ref{fig:fig2}a. The 3D visualization is produced by VAPOR \cite{Li-2019-VAPOR-Atmos..10..488L}.}
    
    \label{fig:fig1}
\end{figure}

We analyze the simulation of an emerging active region (Methods) performed by Chen et al. \cite{ChenF-2022ApJ...937...91C} with the MPS/University of Chicago Radiative MHD (MURaM; \cite{Vogler-2005A&A...429..335V,Rempel-2017ApJ...834...10R}) code. The simulation embodies the emergence of active-region-scale magnetic flux from 10,000\,km beneath to over 100,000\,km above the solar surface. During the 48 h evolution, the emerged magnetic field gives rise to a complex active region and a highly dynamic hot corona. Coronal loops that can be brilliantly seen in the synthetic hot EUV and X-ray passbands form in the active region core as the active region emerges and are sustained till the end of the simulation. We focus on a time period of 6 h 46.8 min starting from 21 h 21.8 min ($t=0$) and a region of $76,800^2$ km$^2$ (marked by the dashed lines in Fig. \ref{fig:fig1}b), which captures the formation of hot coronal loops between two strong sunspots.

Fig. \ref{fig:fig1}a,b shows a snapshot of the 3D view of the synthetic intensity in the 94 \AA\ channel of the Atmospheric Imaging Assembly (AIA; \cite{Lemen-2012SoPh..275...17L}) on board the Solar Dynamics Observatory (SDO; \cite{Pesnell-2012SoPh..275....3P}), a wavelength sampling hot plasma of 6 to 7 MK. One can find abundant high-temperature coronal loops spreading over the complex magnetogram of the photosphere. In the simulation domain, most hot loops are faint or transient. However, this work focuses on the long-lasting bright loops connecting the central pair of polarities P1 and N1. This group of loops reveals interweaving fine structures that map the complex local magnetic fields.

The hot coronal loops in the active region core yield prominent emission in EUV (Fig \ref{fig:fig2}a-c) and SXR wavelengths (Supplementary Video 1) and persist for a time period that is substantially longer than the expected cooling timescale ($\sim$ 0.5 h; Methods). The loops, if viewed as a whole, evolve in a generally steady manner, which is consistent with observations of active region hot loops (Extended Data Fig. 1a,b; \cite{Warren-2011ApJ-six-hours...734...90W}). In a zoom-in view, however, the individual fine structures are seen to evolve dynamically with a tendency of moving downward successively (Supplementary Video 1), which reflects the motion of a series of independently heated ``strands'' (Extended Data Fig. 2, 3d,f). Here and thereafter, the term ``strand'' denotes the fine loop structure identified within this simulation, specifically referring to a quasi-isothermal density structure that experiences a common heating history (Extended Data Fig. 3a-c,e; Methods). We plot the evolution of the differential emission measure (DEM) of the active region core (Fig. \ref{fig:fig2}d), which reveals that a substantial fraction of coronal plasma is maintained at a high temperature of 6 to 10 MK during the whole evolution of loops. This is in accordance with the inversion from observations (Extended Data Fig. 1c). Moreover, since the DEM is summed in each of the small temperature bins spanning $\log T$ = 0.005 (Methods), such a plot exhibits unprecedented details, i.e. numerous slim threads with an average lifetime of $\sim$ 0.5~h, roughly equivalent to the cooling time of such loops (Methods). Each thread is heated to high temperatures of $\gtrsim$ 10 MK rapidly, then gradually cooling down to lower temperatures, without a plateau as shown for the ensemble of loops (except the thread at $\sim$ 2.3~h). This accords with the field-aligned post-heating evolution of the coronal strand \cite{Reale-2014LRSP...11....4R}, implying that the DEM thread corresponds to the elementary loop, which undergoes a single, impulsive heating. By comparison, in observations, because of the large uncertainty in DEM inversion in particular for high temperatures, the independent heating and cooling processes of strands, although widely hypothesized in physical models, have hardly been resolved in practice.

\section{What heats the 10 MK coronal plasma?}\label{sec:heating}

The long-lasting hot loops in the active region core are not heated by the energy deposition along the loops. Instead, they are energized by persistent magnetic reconnections above the loops at a current sheet embedded in a fan-spine-like magnetic topology (Fig \ref{fig:fig1}a,b). The ``fans'' rooted in polarity pairs P1-N2 and P1-N3 reconnect with the ``spine'' rooted in P2-N1. Consequently, the plasma confined by the magnetic fields rooted in P1-N1 is heated and the hot coronal loops are then gradually formed. We present a 2D slice passing through the current sheet and the loop arches (Fig. \ref{fig:fig3_heating_process}a,c) to demonstrate the magnetic and thermodynamic processes involved with magnetic reconnections. The magnetic reconnection occurs within the current sheet (Fig. \ref{fig:fig3_heating_process}b,d,f), which is seen as a slender, sheet-shaped region (thickness of $\sim$ 0.5 Mm). Inflows converge towards the sheet and the magnetic reconnection induces strong outflows (with a velocity of $> 100\ \mathrm{km\ s^{-1}}$) towards the loops; magnetic energy is initially released in the sheet-shaped region with a high volumetric heating rate of $Q_{\rm tot}>10^{-2}\ \mathrm{W\ m^{-3}}$ (Fig. \ref{fig:fig3_heating_process}f; Methods) and then transported to the loops as the magnetic fields contract downward (Fig. \ref{fig:fig3_heating_process}a,c). At a short time interval, super-hot plasmoids are seen to move fast along the current sheet and impact the outflow region (Fig. \ref{fig:fig3_heating_process}b,d; Supplementary Video 2). At a longer time interval, the current sheet could keep existing with continuous converging inflows, maintaining a reconnection rate of the order of 0.1 (Fig. \ref{fig:fig3_heating_process}e). More importantly, so long as the fan-spine-like magnetic topology persists (Extended Data Fig. 4), such magnetic reconnections can on time average provide a continuous energy source to maintain the hot loops for a long time (Extended Data Fig. 5).

Since the current sheet is indeed an extended 3D structure evolving with time (Extended Data Fig. 6), a fixed 2D slice is thus insufficient to reveal its properties. Therefore, we further investigate the heating rate integrated over a finite volume that embodies the current sheet over this time period. First, the volume-integrated heating rate undulates within one order of magnitude of $\sim 10^{19}$ to $10^{20}\ \mathrm{W}$, with an average value of $1.8\times 10^{19}\ \mathrm{W}$ (Fig.~\ref{fig:fig4}a). Such a heating rate seems large and is comparable to the energy release rate of C-class flares \cite{Qiu-2021ApJ...909...99Q}. Nevertheless, considering the large cross section of the loop base (Extended Data Fig. 7), it corresponds to an average heating flux of $1.9\times10^{6}$~W~m$^{-2}$, which is a reasonable value that can sufficiently account for the energy budget of the super-hot 10 MK coronal plasma (Methods). Second, the heating rate is overall steady and persists over a considerably long timescale of $\sim6.8$~h, which is essentially different from the flaring process that usually lasts for only a short period of minutes to tens of minutes. In addition, we also check the dissipation built up by the magnetic braiding, where a myriad of intermittent magnetic energy releases collectively contributes to a steady heating flux along the loop (Methods). In such a simulation, the heating flux is on average $\sim2\times10^5$ W m$^{-2}$ (Extended Data Fig. 8d), which is similar to the results of previous steady heating models \cite{Winebarger-2011ApJ...740....2W}, but one order of magnitude smaller than the threshold required for heating the 10 MK plasma.

More importantly, on small spatial scales, the heating rate manifests as a myriad of impulsive heating events with a duration of $\sim\ 3\ \mathrm{min}$ (Fig. \ref{fig:fig4}c). This is similar to the nanoflare-like heating that is widely employed in one-dimensional coronal loop simulations \cite{Cargill-2015RSPTA-nanoflare.37340260C} and aligns with the hypothesis that ``all coronal heating is impulsive'' \cite{Klimchuk-2015RSPTA.37340256K}. In particular, plasmoids form as a result of locally intense magnetic reconnections, which produce impulsively high heating rates. For instance, the majority proportion of the heating rate originates from a region of a size within 1.5 Mm in the current sheet (Fig. \ref{fig:fig4}c,d). A typical plasmoid forms in the region adjacent to the reconnection site (Fig. \ref{fig:fig4}e), which represents a twisted mini magnetic flux rope in three dimensions (Extended Data Fig. 9). Taking the heating features on both small and large spatial scales into account, we can reach a physical picture that the hot coronal loops are indeed heated by a series of impulsive heating events, but globally manifest a gentle and steady evolution as long as the short-lived heating events occur frequently enough to provide an average heating rate compatible with the total energy loss rate.

We also investigate the 3D kinematics of the super-hot coronal plasma of $> 10$ MK in the active region and find that it is moving downward during the whole period, as shown by the volume-integrated momentum perpendicular to the magnetic field, $mv_{\perp}$ (Fig. \ref{fig:fig4}b). Since $v_\perp$ reflects the motion of magnetic fields in the corona, such long-lasting mass flows reveal the contraction of magnetic fields driven by magnetic tension forces resulting from continuous magnetic reconnections (Fig. \ref{fig:fig3_heating_process}c; \cite{Forbes-1996ApJ...459..330F}).

\section{Driver of the persistent heating}\label{sec:formation}

It is known that magnetic reconnection is the key mechanism responsible for the solar eruptive activities. Remarkably, although a single reconnection event is ephemeral, in growing active regions, continuous magnetic flux emergence can keep magnetic reconnections ongoing.

In our simulation, the positive polarity P1 is the main polarity involved in the fan-spine-like magnetic reconnection during the whole evolution, although it could connect with different negative polarities in the surroundings of P1 as the active region evolves. The magnetic fields connecting P1 reconnect with the ``spine'' rooted in P2-N1 or the open field rooted in N1, thereafter producing the plasma-hosting magnetic fields connecting polarities P1-N1 (Fig. \ref{fig:fig1}; Extended Data Fig. 4). Dynamically, the magnetic reconnections are driven by the vertically rising motion of magnetic fields rooted in P1 and occur at the separatrix layer between the emerging fields and existing fields rooted in N1 (Extended Data Fig. 6), as proposed in the flux-tube tectonics scenario \cite{Priest-2002ApJ-Tectonics...576..533P}. In some cases the emerging magnetic fields can push cold ($\lesssim 10^4$ K) and dense ($\gtrsim  10^{11}\ \mathrm{cm^{-3}}$) plasma from the lower atmosphere to the corona at a height of $\sim$ 20 Mm (Fig \ref{fig:fig5}a,b). The emerging fields then undergo reconnections at the current sheet, resulting in the formation of a coronal loop with a density of approximately $10^{10.5}\ \mathrm{cm^{-3}}$ (Supplementary Video 3). This provides a clear picture of how flux emerges to the corona and gets involved in magnetic reconnections leading to the formation of coronal loops.

In the long term, continuous magnetic reconnections are driven by newly emerging magnetic fields in the active region. Quantitatively, the total positive flux that emerges in the P1 region amounts to $\sim 5\times10^{21}$ Mx during the whole period of 6.8 h (Fig \ref{fig:fig5}d,e; Supplementary Video 4), with an almost steady flux emergence rate of $\sim$ $2\times10^{17}$ Mx s$^{-1}$. From an energy perspective, the emerging magnetic fields transport magnetic energy from the lower to higher atmospheric layers, ultimately releasing it through magnetic reconnections. In particular, as the flux emerges to the current sheet region, the volumetric heating rate has increased fivefold to $\gtrsim 0.25\ \mathrm{ W\ m^{-3} }$ (Fig. \ref{fig:fig5}c), which clearly reveals the transport-release chain. It is the strong and steady flux emergence that can keep the energy chain working for long. At the end of the selected simulation segment, the total unsigned magnetic flux in the loop region reaches $\sim3.4\times10^{22}$ Mx, which is among large active regions with strong hot $\mathrm{Fe\ XVIII}$ emissions in previous statistics \cite{Warren-2012ApJ-survey...759..141W}. Moreover, it is also found in observations that emerging active regions manifest considerably enhanced hot emissions compared to non-emerging ones \cite{Asgari-Targhi-2019ApJ-emergence...881..107A}. It is thus conceivable that the coronal plasma can be heated to very high temperatures in the active region with a strong flux emergence in our simulation.

\section{Discussion}\label{sec:discussion}

We have presented a 3D self-consistent radiative MHD model of hot coronal loops in the cores of an emerging active region, which successfully reproduces the main characteristics of observations including the temperature profile and the long-term steady evolution of the integrated EUV and SXR emissions. Furthermore, our model reveals a range of thermodynamic and radiative details that remote-sensing observations can hardly resolve, e.g. the DEM threads and fine strands. Fundamentally, continuous flux emergence in the active region drives magnetic reconnection that provides the heating energy for the super-hot coronal plasma.

The important finding is that each strand is heated independently and impulsively but the ensemble of strands manifests as a long-lived and stationary loop morphology. For each strand, when magnetic reconnection occurs, it is impulsively heated to temperatures of $\gtrsim 10$ MK and contracts downward, becoming a part of the hot loops. Consequently, the hot loops indeed comprise a series of strands with different temperatures heated with different histories (Extended Data Fig. 8a,b); in addition, the seemingly stable loops are actually very dynamic, in which a stream of newly heated strands moving downward and successively taking over the previously formed ones (Extended Data Fig. 8c). Such a process continues as long as the flux emergence and magnetic reconnection persist, maintaining the apparently stationary morphology of coronal loops as observed. This evolution picture is basically consistent with the one-dimensional multi-stranded coronal loop model \cite{Reale-2014LRSP...11....4R,Klimchuk-2015RSPTA.37340256K}. Furthermore, our simulation advances our knowledge on how the magnetic field drives the energy release process in the three-dimensional space, which reveals the spatio-temporal relationships among the fine structures within coronal loops self-consistently.

From the perspective of remote-sensing observations, in regions with frequent and densely-distributed heating events, considering the complex, veil-like emissivity (Extended Data Fig. 3a) as well as the integration effect, it is challenging to distinguish those fine structures from the overall loop morphology in observed images \cite{Malanushenko-2022ApJ-coronal-veil...927....1M}. Nevertheless, for regions with relatively sporadic heating events, the strand can be clearly identified from the synthetic SXR images (Extended Data Fig. 2a,c). Such a fine emission structure represents the quasi-isothermal super-hot plasma ($\gtrsim10.5\ \mathrm{MK}$) moving with the contracting local magnetic field (Extended Data Fig. 3). Moreover, it is interesting that the width of the super-hot strand in our simulation is measured to be similar to that of warm strands ($1-1.5\ \mathrm{MK}$) resolved by the Hi-C mission (Methods; \cite{Aschwanden-2017ApJ-strand...840....4A,Williams-2020ApJ...902...90W,Williams-2020b-ApJ...892..134W}). Therefore, our simulation provides a theoretical reference for the subsequent high-resolution observations of the high-temperature corona.

Sequential, impulsive energy releases from magnetic reconnections have been extensively reported in particular in flare-like events with a total duration of $\lesssim$ 0.5 h, such as quasi-periodic pulsations in hard X-ray and radio emissions \cite{Zimovets-2021SSRv-QPP-review..217...66Z}, as well as quasi-periodic chromospheric dynamics \cite{Brosius-2018ApJ-Chromospheric-QPPs...867...85B}. Our model of coronal heating, in the short term, is similar to a flare case, both of which comprise a series of short-lived, impulsive energy releases; however, in the long term, they are different in the total duration of heating, which is much longer in the former than that in the latter. Physically, the small impulsive energy releases are accompanied with the plasmoids within the current sheet (Supplementary Movie 5) that are triggered by the tearing mode instability according to relevant theories \cite{Furth-tearing-mode-1963PhFl....6..459F,Pontin-Priest-2022LRSP...19....1P}.

It is known that flux emergence plays a pivotal role in driving solar eruptive events in different scales \cite{Cheung-2014LRSP...11....3C}. For example, some small-scale, persistent UV and EUV brightenings can be powered by interactions between emerging flux and pre-existing magnetic fields \cite{Martinez-Sykora2008ApJ-bright-points...679..871M,Guglielmino-2018ApJ...856..127G,Hansteen-2019A&A...626A..33H}. Our model shows that flux emergence is also the key factor sustaining the intermediate-scale, long-lasting heating of the hottest corona in active regions. In addition to the traditional view that the heating of coronal plasma occurs coincidentally with the upward transport of the energy flux when the emerging magnetic field expands into the corona, our model proposes a new scenario that the flux emergence continuously brings energy to the large current sheet where magnetic reconnection heats the coronal loops that contract downward. This scenario allows for accumulation and release of much more magnetic energy than foot-point braiding and hence provides the strong energy flux that is required to heat the super-hot coronal plasma in the active region core. Likewise, convection and flux emergence are prevalent in solar-like stars and dwarfs \cite{Brun-Browning-2017LRSP-dynamo-review...14....4B}, and thus our model is also promising in explaining the origin of the hottest components of the coronae of these stars \cite{Toriumi-2022ApJS..262...46T}.

\newpage

\begin{figure}[ht]
    \centering
     \includegraphics[width=1.0\textwidth]{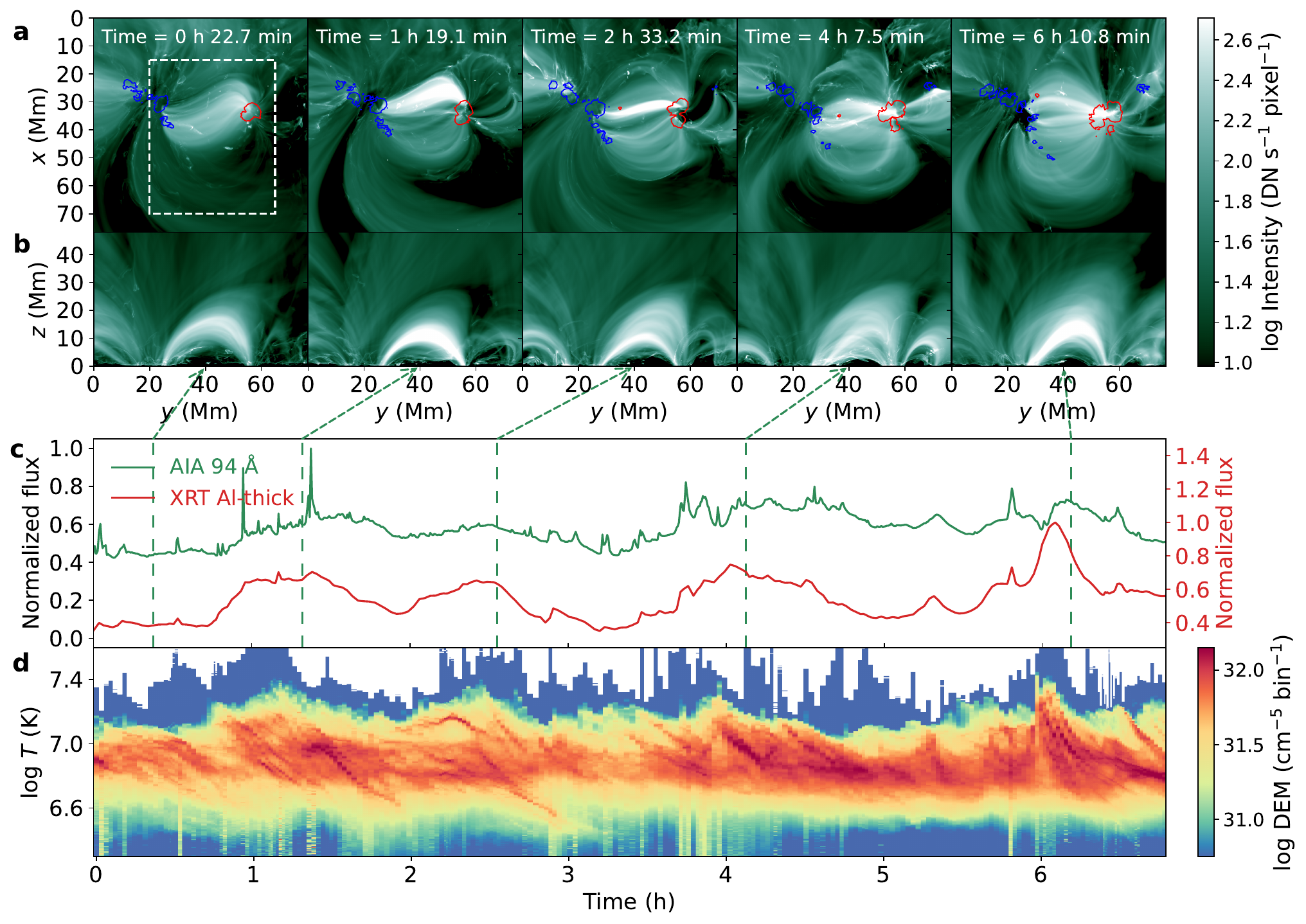}
    \caption{\textbf{An overview of the evolution of the hot coronal loops in active region cores.} \textbf{a,} Synthetic emission in AIA 94 \AA\ integrated along the $z$-direction at five typical times, overlaid with contours of magnetic field $B_z$ of $\pm$ 4000 G. Red and blue colors indicate positive and negative polarities respectively. \textbf{b,} Synthetic emission in AIA 94 \AA\ integrated along the $x$-direction. \textbf{c,} Normalized flux curve in AIA 94 \AA\ and that in XRT Al-thick, an SXR band sampling plasma at temperatures of $\sim$ 10 MK. The flux is calculated by integrating the emission along the $z$-direction over the area outlined by the white dashed box in panel (a). \textbf{d,} Temporal evolution of the plasma DEM in the temperature range of 6.3 $\leqslant$ $\log T$ $\leqslant$ 7.6 with a bin size of $\log T$ = 0.005. The area for integration is the same as that for panel (c).
    }
    \label{fig:fig2}
\end{figure}

\begin{figure}[ht]
    \centering
     \includegraphics[width=1.0\textwidth]{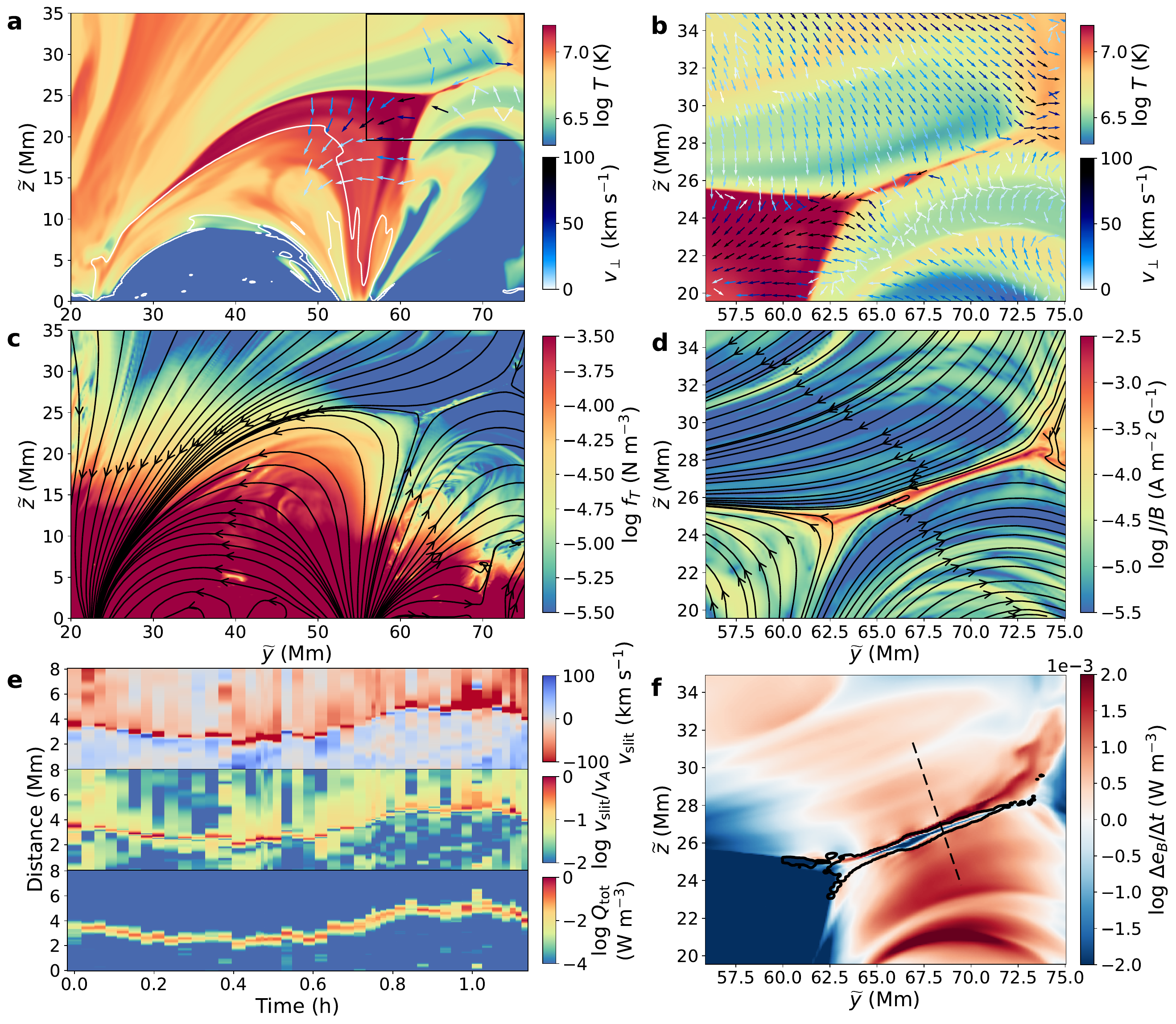}
    \caption{\textbf{A 2D slice passing through the current sheet and hot loops at $\bm{t = 7.2}~\mathbf{min}$ for illustration of the heating mechanism.} \textbf{a,} 2D distribution of temperature (as per the upper colorbar). The white contour outlines the coronal loops with synthetic AIA 94~\AA\ emissivity of $10^{0.5}$~$\mathrm{DN\ s^{-1}\ pixel^{-1}\ Mm^{-1}}$. The arrows represent the velocity vectors perpendicular to the magnetic field, $v_\perp$, projected on the plane, which reflect the motion of magnetic fields. The color of arrows indicates the speed as per the lower colorbar. Detailed properties of the current sheet region (outlined by the black box) are shown in panels (b), (d) and (f). The 2D slice is slightly inclined relative to the $y$--$z$ plane and is thus labeled by $\Tilde{y}$--$\Tilde{z}$. \textbf{b,} A zoom-in view of the panel (a) with the same notations. A movie of panels (a) and (b) is attached in Supplementary Video~2. \textbf{c,} Distribution of magnetic tension force density, $f_T=B^2\mathbf{\hat{b}}\cdot\nabla\mathbf{\hat{b}}/\mu_0$. Black lines represent the magnetic field lines projected on the plane. \textbf{d,} Distribution of the current density normalized by magnetic field strength, $J/B$. \textbf{e,} Time-distance plots of the velocity component projected along the slit, $v_{\rm slit}$, ratio of $v_{\rm slit}$ to the Alfv\'en speed, $v_{\rm slit}/v_A$, and volumetric heating rate, $Q_{\rm tot}$, across the current sheet. Positive (negative) values of $v_{\rm slit}$ refer to upward (downward) motions. The distance is measured from the lower to higher ends of the slit. \textbf{f,} Distribution of the change rate of magnetic energy density, $\Delta e_B/\Delta t$, which is calculated relative to the value measured 203 s before this time. The black contour outlines the region with a volumetric heating rate of $10^{-2}$~W~m$^{-3}$. The black dashed line marks the slit with a length of 8~Mm for the time-distance plots in panel (e).
    }
    \label{fig:fig3_heating_process}
\end{figure}

\begin{figure}[ht]
    \centering
     \includegraphics[width=1.0\textwidth]{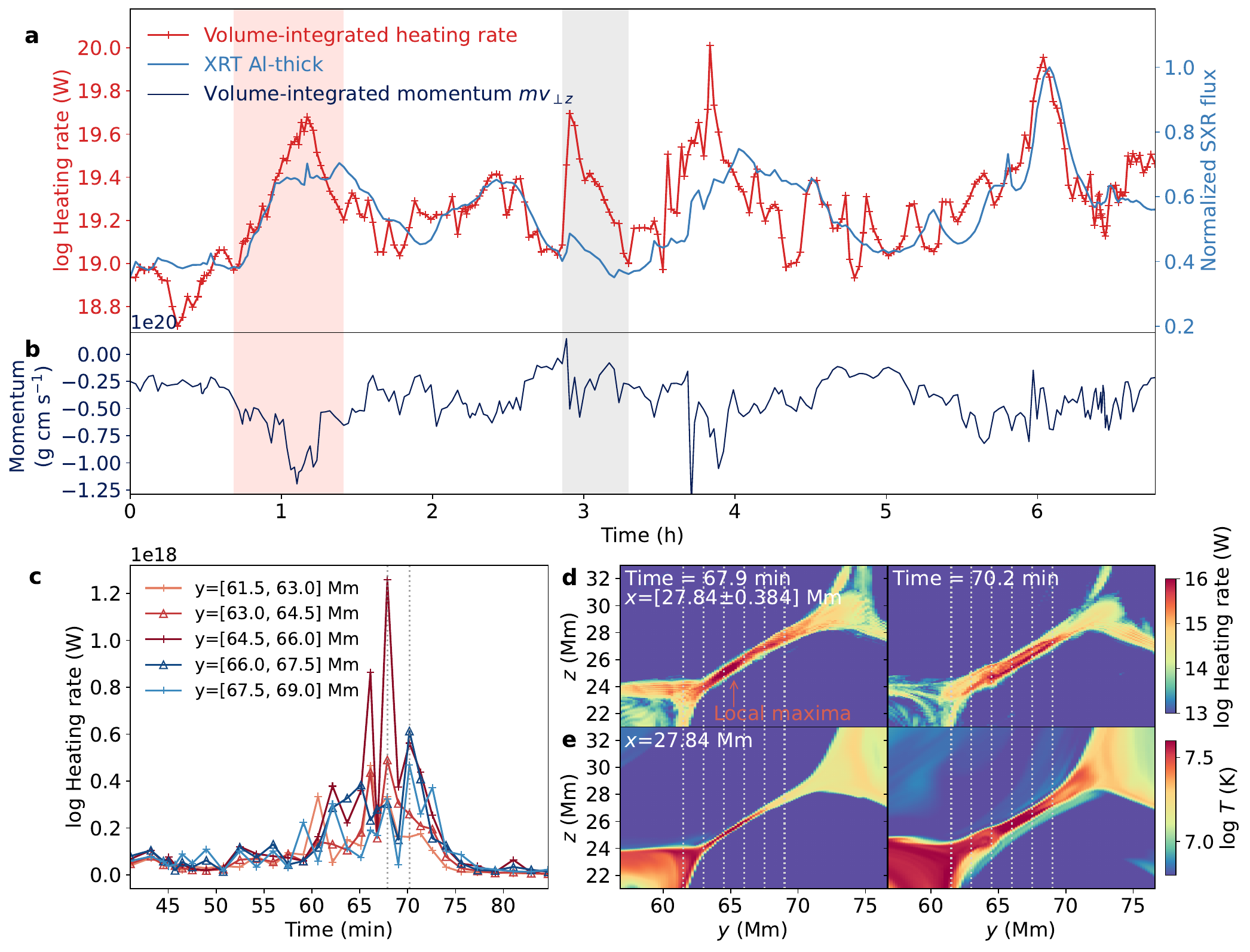}
    \caption{\textbf{Temporal evolution of heating quantities.} \textbf{a,} Evolution of the volume-integrated heating rate (red curve). The 3D volume incorporates the region of magnetic reconnections, which has a domain of $x=[12.5,70]$ Mm, $y=[36.8,76.8]$ Mm, and $z=[21,48]\ \mathrm{Mm}$. The red shaded area marks the time period that is further studied in panel (c). The gray shaded area marks one special magnetic reconnection not detailed in this study, which heats the adjacent small loop at around $y = 70$ Mm resulting from the eruption of twisted magnetic fields. Also plotted is the SXR flux of XRT Al-thick band (light blue curve) which is the same as that in Fig. \ref{fig:fig2}c. \textbf{b,} Evolution of the volume-integrated momentum, $mv_{\perp z}$, i.e. $z$-component of the momentum perpendicular to the magnetic field, from the super-hot plasma of $\gt 10$ MK summed over a 3D box incorporating the region of hot coronal loops. The box has a domain of $x=[15,70]\ \mathrm{Mm}$, $y=[20,65]\ \mathrm{Mm}$, and $z=[0,48]\ \mathrm{Mm}$. \textbf{c,} Evolution of heating rates of five different sub-regions within the current sheet. The sub-regions are marked in panels (d) and (e). \textbf{d,} 2D distribution of the volume-integrated heating rate for a thin slab containing the current sheet. \textbf{e,} 2D distribution of temperature for the plane at $x=27.84$ Mm.
    }
    \label{fig:fig4}
\end{figure}

\begin{figure}[ht]
    \centering
     \includegraphics[width=1.0\textwidth]{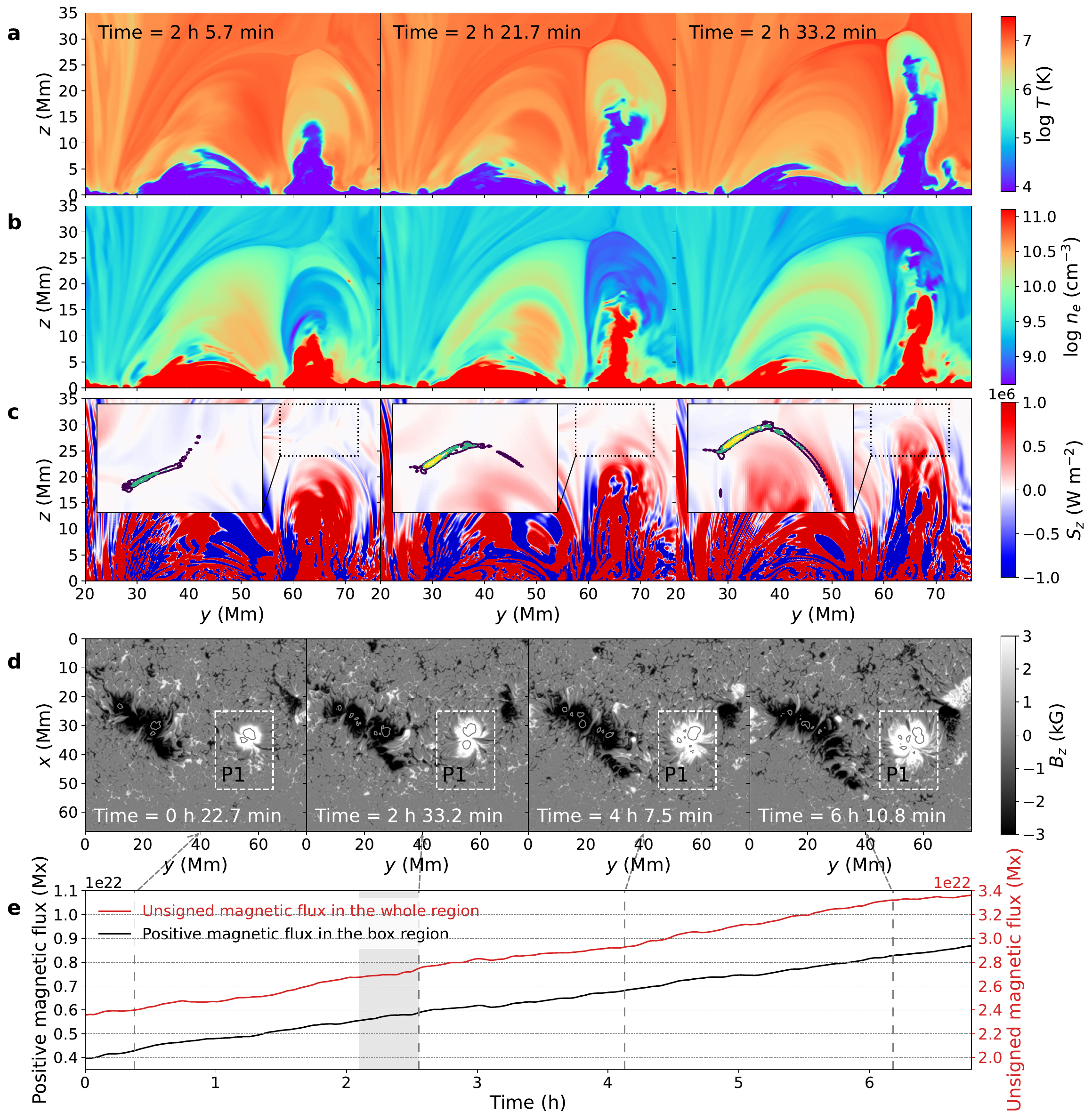}
    \caption{\textbf{Flux emergence in the active region.} \textbf{a,} 2D distribution of temperature at $x = 47.5$ Mm for three typical time instants. \textbf{b,} 2D distribution of the electron number density. \textbf{c,} 2D distribution of the $z$-component of the Poynting flux, $S_z$. The solid rectangular frame presents a zoom-in view of $S_z$ of the current sheet region (outlined by the dotted box), where contours of volumetric heating rate, $Q_{\rm tot}$, are overlaid with levels of $0.01\ \mathrm{W\ m^{-3}}$ (violet), $0.05\ \mathrm{W\ m^{-3}}$ (green) and $0.25\ \mathrm{W\ m^{-3}}$ (yellow). A movie of panels (a)--(c) is attached in Supplementary Video~3. \textbf{d,} Magnetograms ($B_z$) of the photosphere at four typical time instants, overlaid with gray contours of $B_z$ at levels of $\pm$ 6000 G. A movie of panel (d) is attached in Supplementary Video~4. \textbf{e,} Temporal variation of the magnetic flux. The red line refers to the unsigned magnetic flux summed over the whole region shown in panel (d). The black line refers to the positive magnetic flux summed over the white dashed box region in panel (d). The shaded area marks the duration of flux emergence shown in panels (a)--(c).
    }
    \label{fig:fig5}
\end{figure}

\clearpage
\newpage

\section{Methods}\label{sec:method}

\subsection{Simulation Setup}
The simulation covers a Cartesian domain of $L_x \times L_y \times L_z = 196.608$ Mm $\times 196.608$ Mm$ \times 122.88$ Mm resolved by grids of $N_x \times N_y \times N_z = 1024 \times 1024 \times 1920$, with grid spacings of $\Delta x = 192$ km, $\Delta y = 192$ km, and $\Delta z = 64$ km. The photosphere is defined as the $z = 0$ Mm surface where the average optical depth equals unity, beneath which is the convective zone extending downward for 9.6 Mm. 

The initial condition for the active region simulation adapts a snapshot of a quiet Sun simulation with the same grid spacings. The quiet Sun corona, which has been evolved for a sufficiently long time to achieve a dynamic equilibrium, is maintained by the magneto-convection braiding magnetic field lines. Building upon this, the formation and evolution of the active region are driven by the emerging magnetic flux from the uppermost layer of the convective zone in a global solar convective dynamo simulation \cite{FanYH-2014ApJ...789...35F}. A time series of magnetic and velocity fields that embody the emergence of super-equipartition flux bundles in the dynamo simulation is implemented as a time-dependent bottom boundary of the active region simulation. In this way, we provide a physical description of the entire process encompassing the generation and transport of magnetic flux from the deep solar interior to the corona. The coupling between the bottom boundary and the MURaM simulation is described by Chen et al. \cite{ChenF-2017ApJ...846..149C}. The comprehensive description of the simulation setup can be referred to Chen et al. \cite{ChenF-2022ApJ...937...91C}.

At the top boundary, magnetic fields are the extrapolated potential fields and hydrodynamic variables including the horizontal flows follow the symmetric boundary condition; the vertical flows are allowed to leave but they are strongly damped at ghost cells for numerical stability. The lateral boundary conditions are periodic. The damping of outflows at the top boundary and periodic lateral boundary lead to a trapping of hot plasma in the high corona ($z>60$\,Mm). We perform a control experiment, where the outflows at the top boundary are not damped. The open boundary allows the hot plasma accumulated near the top boundary to escape and reduces the emission measure in $\log T>6.5$ by about a factor of 5. Thus, the mean coronal temperature slightly decreases to 3.7\,MK. The hot loops in the active region core discussed in the paper are in the lower half of the domain and thus not affected by the top boundary.

\subsection{Heating rate}

The volumetric heating rate considered in this work (Fig. \ref{fig:fig3_heating_process}, \ref{fig:fig4}) is the sum of numerical viscous heating and resistive heating, $Q_{\rm tot} = Q_{\rm vis} + Q_{\rm res}$. The quantity $Q_{\rm vis}$ ($Q_{\rm res}$) measures the dissipation of kinetic (magnetic) energy by the numerical scheme described in Rempel et al. \cite{Rempel:2014}. For reference, the magnetic Reynolds number at the scale of the simulation region is of the order of $10^4$. During the calculation, $Q_{\rm vis}$ is implicitly added to the internal energy because the code solves the conservation of plasma energy, i.e., the sum of the internal and kinetic energy of the plasma, whereas $Q_{\rm res}$ is estimated by the formula given in Rempel et al. \cite{Rempel-2017ApJ...834...10R} and added to the energy equation as a source term. Both quantities are stored at the same output cadence as that of other 3D physical quantities to facilitate analysis as shown in this paper. In this way, not only the amount of the heating rate but also its variation in time and space are determined spontaneously through the evolution. This ensures a high degree of self-consistency in calculating the heating rate in our simulation.

In this simulation, the granular motions in the photosphere braid magnetic field lines \cite{Breu-2022A&A...658A..45B}. The braiding provides a net upward energy flux into the corona, where the energy is released by the numerical viscous and resistive dissipation. The distribution of $Q_{\rm vis}$ and $Q_{\rm res}$ is very intermittent in both space and time \cite{Rempel-2017ApJ...834...10R}, and the volume-integrated heating rate consists of a stable mean level superimposed by fluctuations on all time scales \cite{ChenF-2022ApJ...937...91C}. In general, the process of energy transport and dissipation is a numerical realization of the magnetic braiding scenario proposed by Parker \cite{Parker-1983ApJ...264..642P} and consistent with what has been found in previous realistic 3D coronal models \cite{Gudiksen+Nordlund:2005a,Bingert-2011A&A...530A.112B,Hansteen-bifrost-2015ApJ...811..106H,Breu-2022A&A...658A..45B}. Note that the magnetic braiding provides a steady heating flux to generally sustain the temperature of the solar corona. However, this heating flux is insufficient to heat the super-hot corona found in the cores of active regions.

We also calculate the heating rate per particle, $Q_{\rm tot}/n = Q_{\rm tot}/(n_{e}+n_{H}+n_{He})$, which more directly reflects the rate of temperature change in response to heating. As shown in Extended Data Fig. 6, the heating rate per particle is mainly concentrated in the coronal regions where the magnetic connectivity dramatically changes. In these regions, the current density is high and the coronal plasma is heated by magnetic reconnections.

\subsection{Observational data}
We provide observational analysis of the long-lasting hot coronal loops in the cores of the active region NOAA 12242 located at
S20$^{\circ}$E25$^{\circ}$ on 2014 December 15 (Extended Data Fig. 1). Note that as a theory-based model, we aim to describe a more generic scenario rather than exactly reproduce a specific active region. Therefore, the purpose of including one observational example is to offer the overall thermal and radiative characteristics that can be used to compare with synthetic remote-sensing results from our simulation.

The loops undergo a relatively steady evolution for more than 15 h, and the data shown here are taken from 10:30 UT to 19:30 UT. In Extended Data Fig. 1a, we show the EUV 94 \AA\ images of the hot loops as observed by the AIA on board the SDO, which has a pixel size of 0.6$''$ and a spatial resolution of 1.5$''$ \cite{Lemen-2012SoPh..275...17L}. The AIA 94~\AA\ emission is mainly from the $\mathrm{Fe\ XVIII}$ line with an effective temperature of $\log T \sim$ 6.8. The light curve of the Geostationary Operational Environmental Satellite (GOES) soft X-ray (SXR) flux and the normalized AIA 94~\AA\ intensity are shown in Extended Data Fig. 1b, both of which reflect the thermal emission of hot plasma. Although the SXR flux maintains a high magnitude comparable to C1- or C2-class flares, considering its steady evolution trend, no flare is recorded during the nine hours. Moreover, we inverse the emission measure (EM) of the selected region marked by the red rectangle in Extended Data Fig. 1a making use of observations of AIA 94~\AA, 131\ \AA, 171\ \AA, 193\ \AA, 211\ \AA, and 335\ \AA\ bands at a temporal cadence of 1 min. The EM is calculated with the method provided by Su et al. \cite{Su2018}, which outputs the distribution of plasma density squared in the temperature range of 5.5 $\leqslant$ $\log T$ $\leqslant$ 7.6 with a bin size of $\log T$ = 0.05. Here we only focus on the plasma with hot emission in the temperature range of 6.0 $\leqslant$ $\log T$ $\leqslant$ 7.2.

\subsection{Numerical DEM analysis}

DEM analysis reveals the radiative intensity contributions from different temperature components. In observations, this is accomplished by synthesizing the radiative intensities from different wavelength bands and performing inversions (Extended Data Fig. 1c). In numerical simulations, with access to temperature and number density of all grids in the three-dimensional space, we can directly calculate the DEM distribution along the z-axis by

\begin{equation}
    \mathrm{DEM}(x,y,T)=\int n_e^2(x,y,z,T)M(T)\mathrm{d}z.
\end{equation}
Since the DEM is calculated within a finite temperature bin of $\Delta\log T=0.005$, the function $M(T)$ serves as a masking function, taking a value of 1 when the temperature is within the given temperature bin and 0 otherwise. DEMs of 260 temperature bins in the range of $6.3 \leqslant \log T \leqslant 7.6$ are calculated and shown in Fig. \ref{fig:fig2}d. Considering the limited grid number, we have tested using even smaller temperature bins; however, this does not lead to finer details in the time-DEM map.

\subsection{Energy flow}

Thermal conduction and radiation are the two primary cooling mechanisms of the solar corona in active regions. Quantitatively, for the corona with a temperature of $T_0\sim2.5$ MK, energy losses of the two mechanisms are $F_c\ \sim 10^2-10^4$ $\mathrm{W\ m^{-2}}$ and $F_r\ \sim 5\times10^3$ $\mathrm{W\ m^{-2}}$ respectively, which result in an overall energy loss of $F_0\sim10^4$ $\mathrm{W\ m^{-2}}$ \cite{Withbroe-Noyes-1977ARA&A..15..363W}. Taking this value as a reference, the combined cooling energy flux of the corona with higher temperatures can be estimated by the scaling law \cite{Rosner-RTV-1978ApJ...220..643R}

\begin{equation}
    F(T)\propto(T/T_0)^\frac{7}{2}F_0.
\end{equation}
In our case, the coronal plasma can be heated to over 10 MK, and we obtain an extremely large energy loss flux of $F(\mathrm{10\ MK})\sim1.28\times10^6\ \mathrm{W\ m^{-2}}$. Therefore, a feasible heating mechanism has to provide a roughly equivalent energy flow to balance the energy loss. The heating mechanism that we propose in this work successfully provides and maintains a high-level heating rate summed over a 3D volume embodying the current sheet during the whole segment of the simulation (Fig. \ref{fig:fig4}). To more directly compare it with the energy loss flux, we take a horizontal slice across the loop near the footpoints at $z=0.256\ \mathrm{Mm}$ and calculate the cross-sectional area of the loop to be $9.55\times10^{16}\ \mathrm{cm^2}$ (Extended Data Fig. 7). Dividing the volume-integrated heating rate by the area, we find that the mean heating rate of $\sim1.82\times10^{19}\ \mathrm{W}$ corresponds to an energy flux of $\sim1.90\times10^6\ \mathrm{W\ m^{-2}}$, which is larger than the energy loss flux of the coronal plasma at $10$ MK. Consequently, we conclude that the heating mechanism in this work is feasible to heat the solar corona in the cores of the active region to $10$ MK and even higher. 

In addition, we also integrate the volumetric heating rate along the ten magnetic field lines (Extended Data Fig. 8d) and obtain an average heating flux of $\sum_{i=1}^{10}(\int_{L_i} Q_{\rm tot} dl)/10\sim2.03\times10^5$ W m$^{-2}$, which represents the heating flux provided by the magnetic braiding. However, it is one order of magnitude smaller than the energy budget for heating the 10 MK coronal plasma in active region cores.

\subsection{Cooling timescale of the coronal loop}
One essential conclusion of this work is that the super-hot coronal plasma can persist for a long time ($\sim$ 6.8 h), which is, however, heated by a series of short-lived impulsive magnetic reconnections ($\sim$ 3 min). Here, the nomenclature of ``long'' and ``short'' is defined by comparing them with the cooling time of the super-hot coronal loops. From theoretical perspective, the cooling timescale can be estimated by applying the equation  \cite{Cargill-1995ApJ...439.1034C}

\begin{equation}
    \tau_{\mathrm{cool}} = 2.35\times10^{-2}\ L^{\frac{5}{6}}T_0^{-\frac{1}{6}}n_0^{-\frac{1}{6}},
\end{equation}
where $L$ is the half-length of the coronal loop; $T_0$ and $n_0$ are the initial temperature and number density when the loop cooling starts. According to the quantitative properties of the loops (Extended Data Fig. 8), we take typical values of $L=23$ Mm, $T_0=10$ MK, and $n_0=10^{10.4}\ \mathrm{cm^{-3}}$. As a result, we obtain a typical cooling time of $\tau_{\mathrm{cool}}=31$ min, which aligns with the cooling timescale of $\sim$ 0.5 h as revealed in the time-DEM maps (Fig. \ref{fig:fig2}d). Note that this equation can only be applied to estimate the cooling time of the high-temperature coronal loop, which cools initially through conduction and then through radiation. Our case matches the precondition.

\subsection{Width of the coronal strand}
We measure the width of one typical super-hot coronal strand based on two different methods. The first method follows the routine of observational analysis by the Hi-C group \cite{Williams-2020ApJ...902...90W}. By fitting the synthetic XRT Al-thick intensity profile (Extended Data Fig. 2a) using a Gaussian function, we obtain a strand width (FWHM) of $\sim 635\ \mathrm{km}$ (Extended Data Fig. 2b). The second method makes full use of the 3D information of the simulation data. We determine the cross-sectional profile of the super-hot strand with a temperature of $T>10.5\ \mathrm{MK}$ within an $x-z$ plane, which allows to locate the seed points at its emission center and along its periphery (Extended Data Fig. 3a-d). The magnetic field lines passing through these seed points spatially align with the strand identified in the synthetic XRT image (Extended Data Fig. 2c). Given this, we quantify the strand width by calculating the mean distance between the center magnetic field line and the edge field lines, resulting in a strand width of $\sim 200-700\ {\rm km}$ (Extended Data Fig. 2d). The width measurements obtained from both methods converge at similar values. Interestingly, when comparing with observational results, we find that despite the temperature of the strand increasing tenfold, the width of the super-hot strand is measured to be similar to that of warm strands (1-1.5 MK) resolved by the Hi-C mission, which is approximately 500 km \cite{Aschwanden-2017ApJ-strand...840....4A,Williams-2020ApJ...902...90W,Williams-2020b-ApJ...892..134W}. As noted earlier, the term ``strand'' specifically refers to the fine loop structure identified in this simulation through radiative and thermodynamic characteristics. This serves as a reference for observations based on the current simulation parameters. Nevertheless, considering the relatively small magnetic Reynolds number in the simulation compared to the real Sun and the potential impact of the numerical resolution \cite{Rempel:2014,Rempel-2017ApJ...834...10R}, further investigations are required to establish its precise correspondence with the finest component of coronal loops in the actual solar environment.
\backmatter

\bmhead{Data availability}
The simulation data and analysis tools are available at the Solar Data Center of Nanjing University 
(\href{https://sdc.nju.edu.cn/d/34ec4acecf294c21be11/}{https://sdc.nju.edu.cn/d/34ec4acecf294c21be11/}).

\bmhead{Code availability}
We have opted not to make the MURaM code publicly available since it is under frequent updating, and running the code needs expert assistance. The numerical methods used in the code are provided in refs \cite{Vogler-2005A&A...429..335V,Rempel:2014,Rempel-2017ApJ...834...10R}. Interested readers are invited to contact the corresponding authors for more information.





\bmhead{Acknowledgments}







We thank Jinhan Guo for help in data processing and Zhuofei Li, Jun Chen, Yulei Wang and P. F. Chen for their valuable discussions. M.D.D. and F.C. are supported by National Key R\&D Program of China under grants 2021YFA1600504 and 2022YFF0503004 and by NSFC under grants 12373054,  12127901 and 12333009. F.C. is also supported by the Program for Innovative Talents and Entrepreneurs of Jiangsu Province. Z.L. is supported by the Postgraduate Research \& Practice Innovation Program of Jiangsu Province under grant KYCX22\_0107. C.W. is supported by Postgraduate Research \& Practice Innovation Program of Jiangsu Province KYCX23\_0118. The simulation data for analysis are based upon the work supported by the National Center for Atmospheric Research, which is a major facility sponsored by the National Science Foundation under Cooperative Agreement No. 1852977. The high-performance computing support is offered by Cheyenne (doi:10.5065/D6RX99HX) provided by NCAR’s Computational and Information Systems Laboratory, sponsored by the National Science Foundation.

\bmhead{Author contributions}
Z.L. analyzed the simulation data. F.C. conceived the study and provided the data. M.D.D. supervised the project. C.W. analyzed the observational data. Z.L., F.C. and M.D.D. wrote the manuscript. C.W., Y.D. and X.C. joined discussions and contributed to the revision of the manuscript.

\bmhead{Competing interests}
The authors declare no competing interests.

\bmhead{Captions for supplementary videos}
\begin{itemize}
\item Supplementary Video 1. Temporal evolution of the hot coronal loops in active region cores. Left panels are the synthetic emission in AIA 94~$\mathrm{\AA}$. Right panels are the synthetic emission in XRT Al-thick. The notations are the same as those of Fig. \ref{fig:fig2}a.
\item Supplementary Video 2. Animation for Fig. \ref{fig:fig3_heating_process}a,b.
\item Supplementary Video 3. Animation for Fig. \ref{fig:fig5}a-c.
\item Supplementary Video 4. Animation for Fig. \ref{fig:fig5}d.
\item Supplementary Video 5. Temporal evolution of the current sheet region at $x=27.648$ Mm with a higher temporal resolution. The left panel displays the 2D distribution of temperature. The right panel presents the current density normalized by magnetic field strength, $J/B$.
\end{itemize}








\newpage


\setcounter{figure}{0}

\begin{figure}[h]
    \centering
    \includegraphics[width=1.0\textwidth]{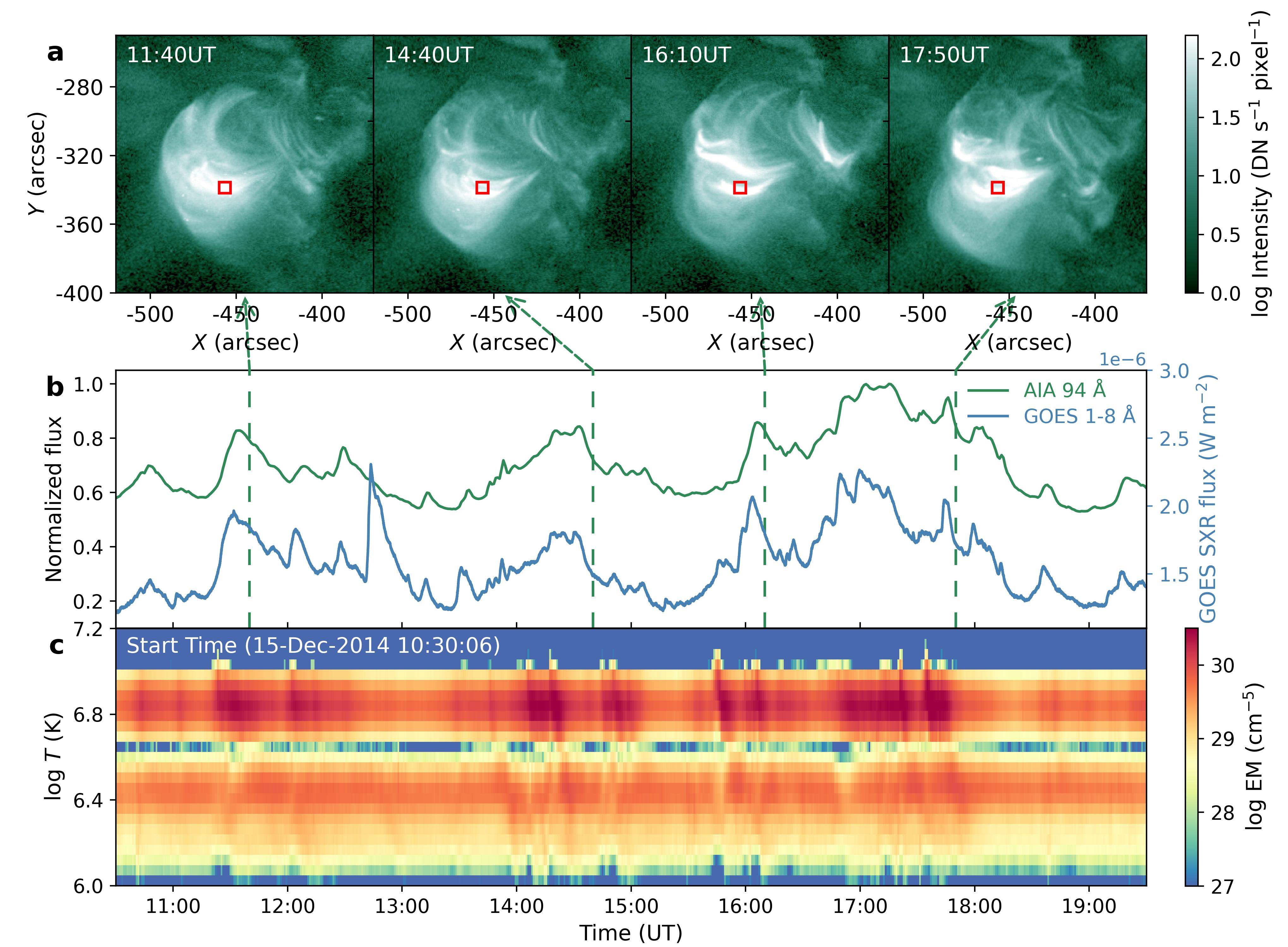}
    \captionsetup{labelformat=extended}
    \caption{\textbf{Evolution of the long-lasting hot coronal loops in the cores of the active region NOAA 12242 observed by SDO/AIA.} \textbf{a,} Emission in AIA 94 $\mathrm{\AA}$ at four snapshots. \textbf{b,} Normalized flux curve in AIA 94 $\mathrm{\AA}$ integrated within the whole region shown in panel (a). The blue curve represents the $1-8\ \mathrm{\AA}$ soft X-ray flux observed by the Geostationary Operational Environmental Satellite (GOES). \textbf{c,} Temporal evolution of emission measure (EM) in the region outlined by the red rectangle in panel (a).}
    \phantomsection
    \label{efig:observation}
\end{figure}

\begin{figure}[h]
    \centering
    \includegraphics[width=1.0\textwidth]{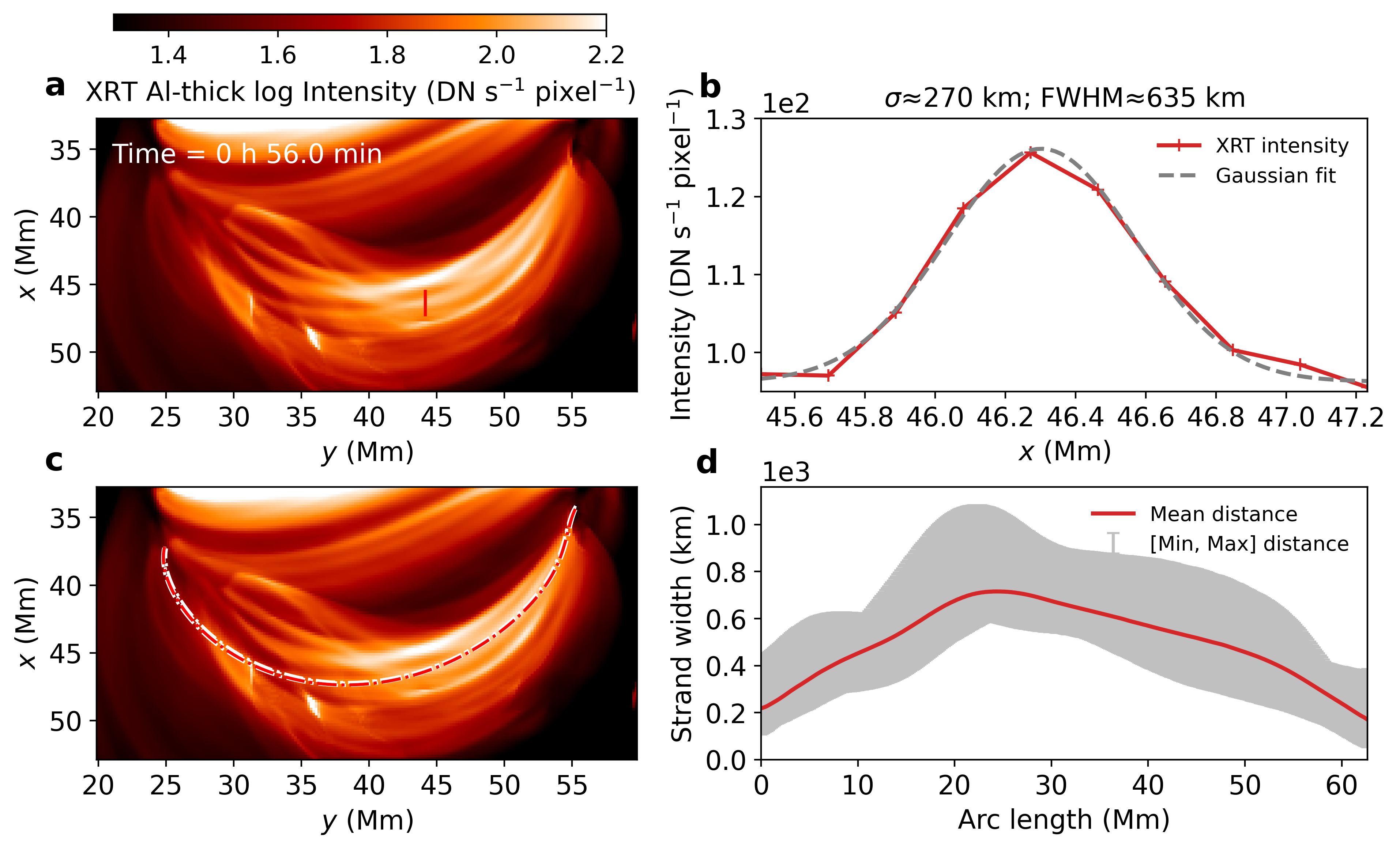}
    \captionsetup{labelformat=extended}
    \caption{\textbf{Identification of the coronal strand and measurements of its width.} \textbf{a,} A zoom-in view of the synthetic SXR emission in the XRT Al-thick band. \textbf{b,} Distribution of the XRT intensity across the coronal strand. Shown here is the intensity along the red short line in panel (a). The grey dashed line represents the Gaussian fit of the XRT intensity. A constant term is added to the Gaussian function to account for the background intensity. The fitting yields the standard deviation, $\sigma$, and the full width at half maximum (FWHM) of the intensity profile. \textbf{c,} Projection of the magnetic field lines along the strand over the SXR emission. The colors of the lines correspond to the colors of seed points in Extended Data Fig. 3b-d, where the red line traces the emission center of the strand. \textbf{d,} Measurement of strand width along the arc length. The width is calculated as twice the mean distance between the central magnetic field line (red) and the edge field lines (white). The grey area shows the range of strand width defined as twice the minimum and maximum distances between the central and edge lines.}
    \phantomsection
    \label{efig:strand01}
\end{figure}

\begin{figure}[h]
    \centering
    \includegraphics[width=1.0\textwidth]{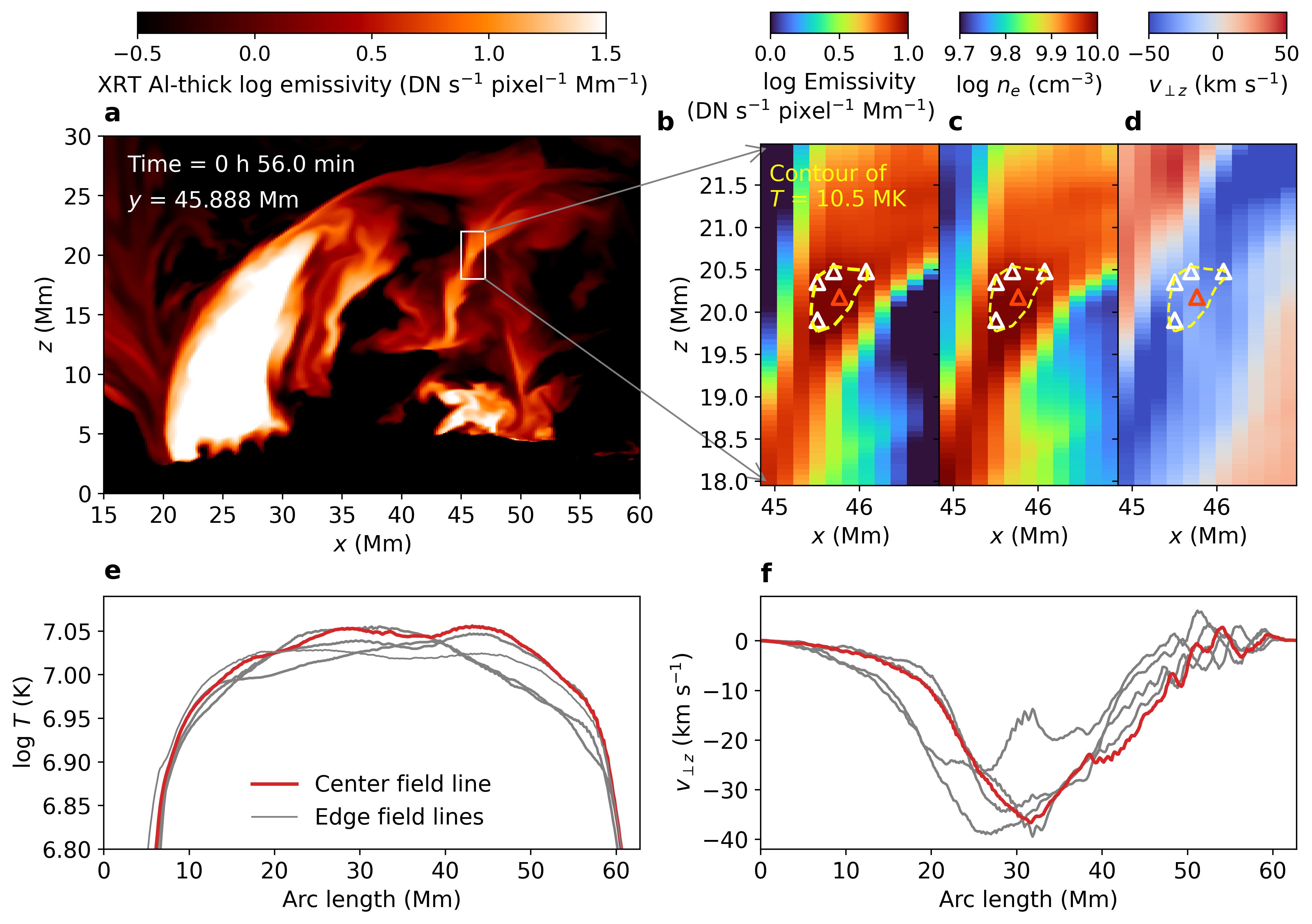}
    \captionsetup{labelformat=extended}
    \caption{\textbf{Thermodynamic properties of the coronal strand.} \textbf{a,} Synthetic SXR emissivity in the XRT Al-thick band at the $y=45.888$ Mm plane. The white box highlights the coronal strand that is identified in Extended Data Fig. 2. \textbf{b,} A zoom-in view of the XRT emissivity. The yellow dashed contour outlines the coronal strand with a temperature of $10.5$ MK. The white triangles represent the seed points at the edge of the strand; the red triangle represents the seed point at the emission center of the contoured strand region. \textbf{c,} Number density, $n_e$, of the zoom-in region. \textbf{d,} $z$-component of the velocity perpendicular to the magnetic field, $v_{\perp z}$, of the zoom-in region. \textbf{e,} Distribution of temperature along the magnetic field lines integrated through the center seed point (red curve) and edge seed points (grey curves) as denoted in panels (b)--(d). \textbf{f,} Distribution of $v_{\perp z}$ along the center and edge magnetic field lines, with the same notations as in panel (e).}
    \phantomsection
    \label{efig:strand02}
\end{figure}

\begin{figure}[h]
    \centering
    \includegraphics[width=1.0\textwidth]{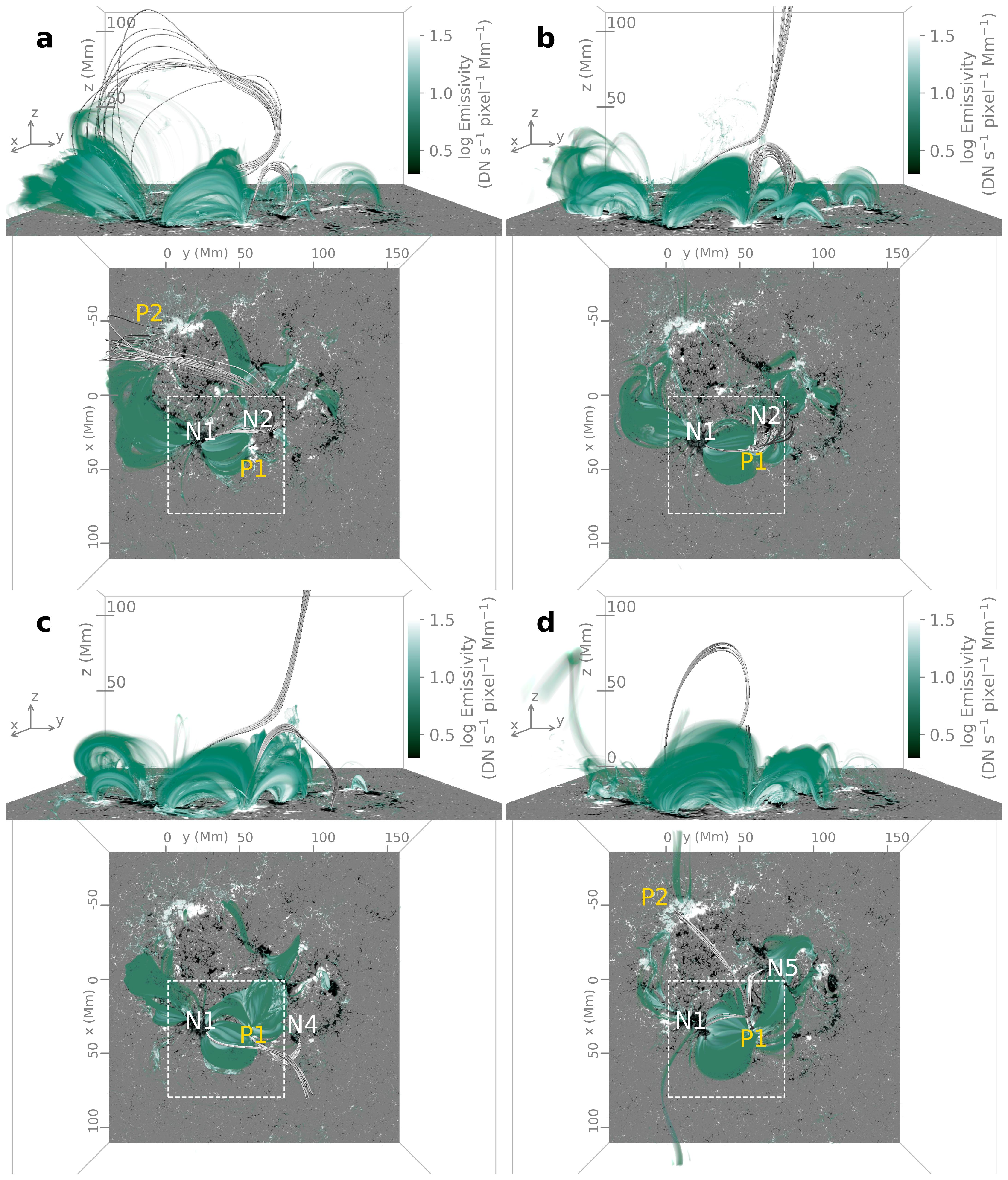}
    \captionsetup{labelformat=extended}
    \caption{\textbf{3D modelling of EUV coronal loops in active region cores.} Four snapshots at $t = 56.0$~min (panel a), 2~h~33.2~min (panel b), 3~h~53.4~min (panel c) and 6~h~3.4~min (panel d) are shown. The notations are the same as those of Fig. \ref{fig:fig1}.}
    \phantomsection
    \label{efig:vapor_3D}
\end{figure}

\begin{figure}[h]
    \centering
    \includegraphics[width=1.0\textwidth]{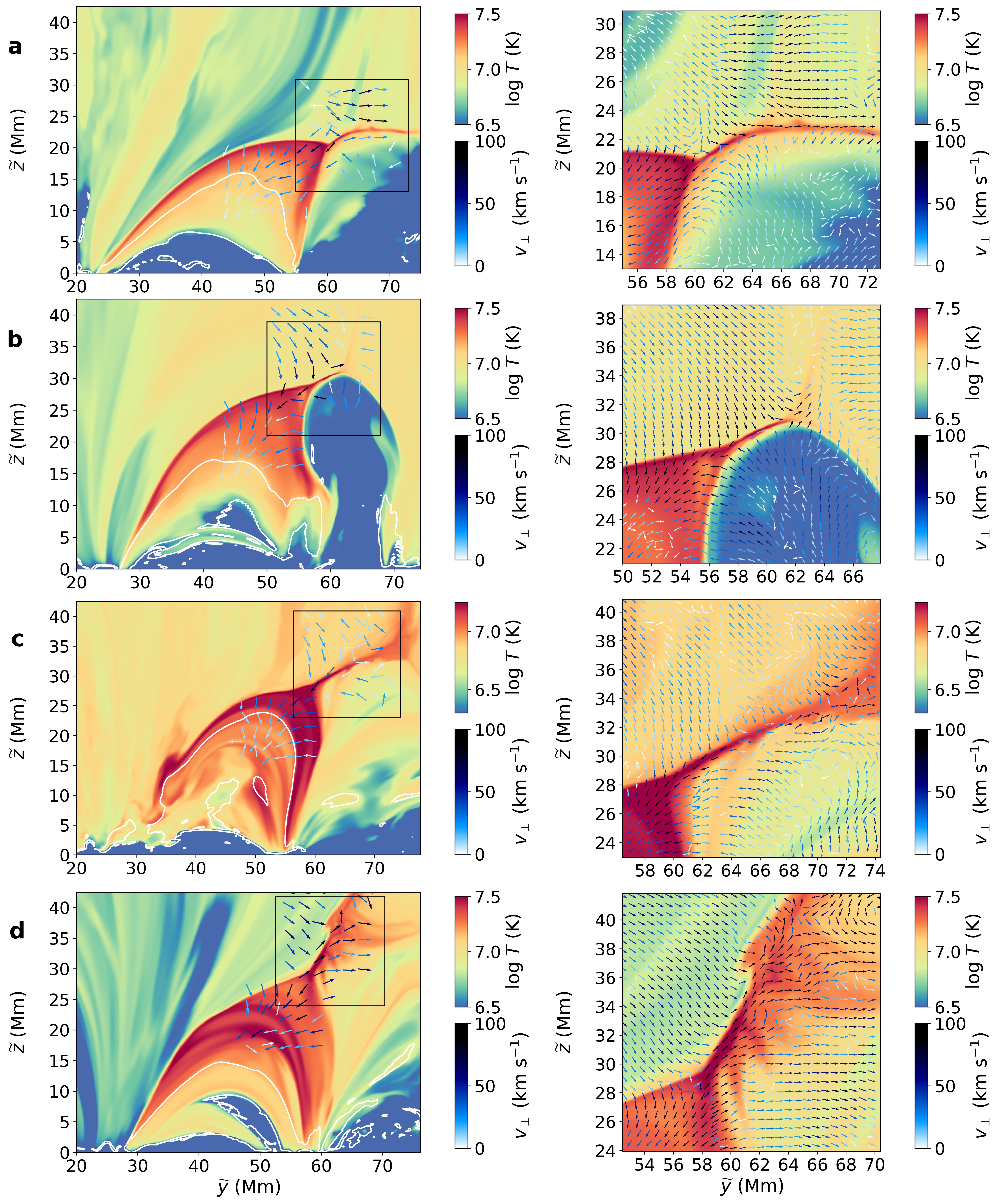}
    \captionsetup{labelformat=extended}
    \caption{\textbf{2D slices passing through the current sheet and hot loops.} Four snapshots at $t = 56.0$~min (panel a), 2~h~33.2~min (panel b), 3~h~53.4~min (panel c) and 6~h~3.4~min (panel d) are shown. The notations are the same as those in Fig. \ref{fig:fig3_heating_process}a,b.}
    \phantomsection
    \label{efig:four_heating2d}
\end{figure}

\begin{figure}[h]
    \centering
    \includegraphics[width=1.0\textwidth]{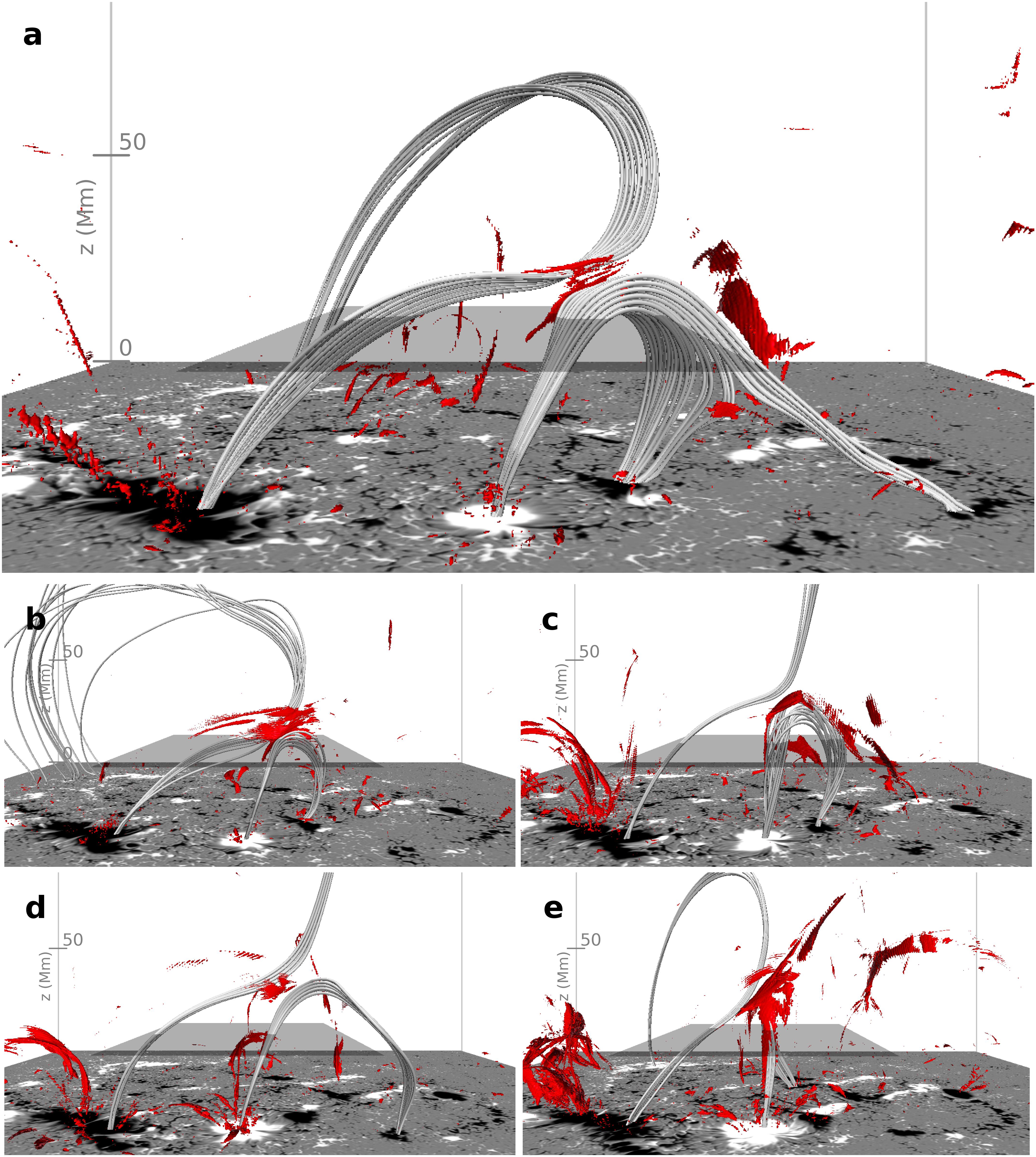}
    \captionsetup{labelformat=extended}
    \caption{\textbf{3D contours of the heating rate per particle, $Q_{\rm tot}/n$, in the active region.} Five snapshots at $t =$~7.2~min (panel a), 56.0~min (panel b), 2~h~33.2~min (panel c), 3~h~53.4~min (panel d) and 6~h~3.4~min (panel e) are shown. The contour level is $10^{-17}\ \mathrm{W}$. The white solid tubes represent the magnetic fields the same as those in Fig. \ref{fig:fig1} and Extended Data Fig. 4. The grey slice represents the bottom surface of the domain used for integrating the heating rate in Fig. \ref{fig:fig4}a.}
    \phantomsection
    \label{efig:Qtotpp_vapor}
\end{figure}

\begin{figure}[h]
    \centering
    \includegraphics[width=1.0\textwidth]{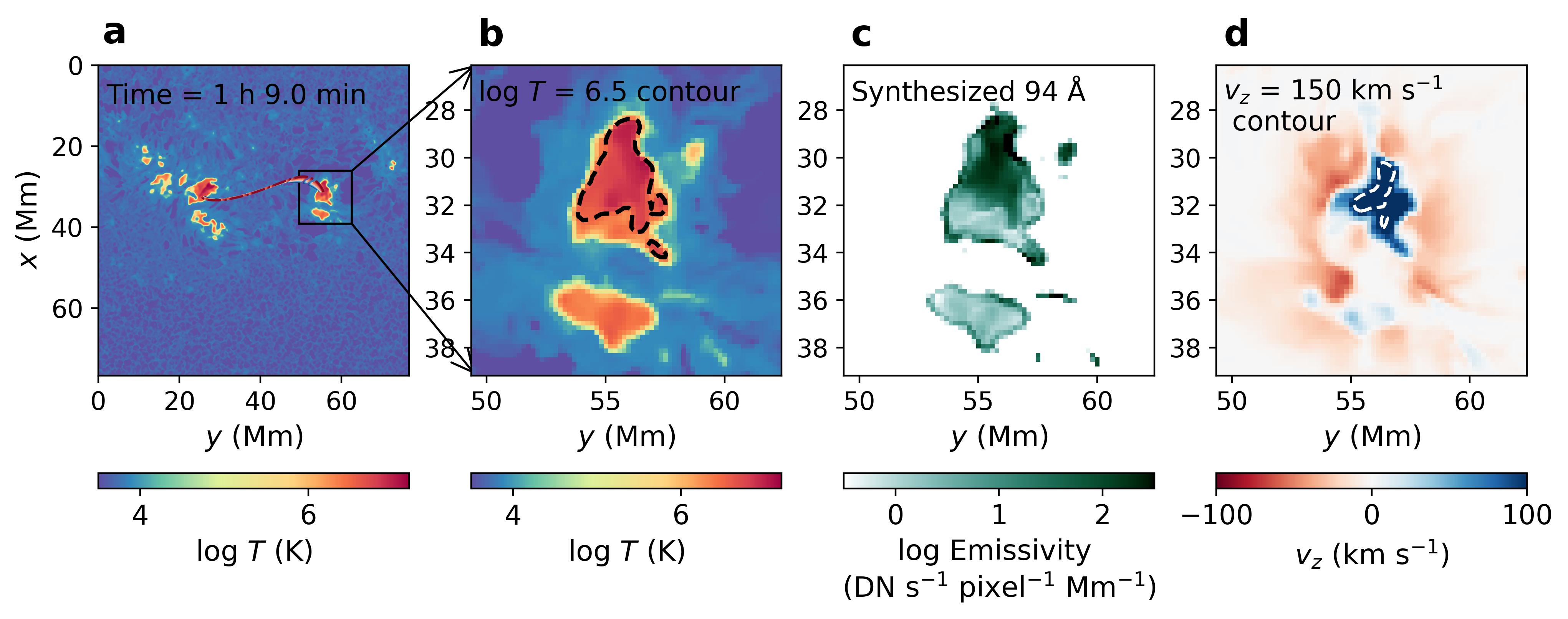}
    \captionsetup{labelformat=extended}
    \caption{\textbf{Quantities in the cross section of the hot coronal loops at their feet.} \textbf{a,} 2D distribution of temperature at $z$ = 0.256 Mm. The colored dash-dotted lines outline the ten magnetic field lines passing through the seed points in Extended Data Fig. 8a. \textbf{b,} A zoom-in view of the footpoint region of the loops. The black dashed contour outlines the cross section of the hot loops (log $T>6.5$), with an area of $9.55\times10^{16}$ cm$^2$. \textbf{c,} Synthetic AIA 94~$\mathrm{\AA}$ emission at the footpoint region. The colorbar is reversed for better visualization. \textbf{d,} Velocity component $v_z$ at the footpoint region. The white dashed line outlines the contour of $v_z=150$ km s$^{-1}$.}
    \phantomsection
    \label{efig:loop_footpoint}
\end{figure}

\begin{figure}[h]
    \centering
    \includegraphics[width=1.0\textwidth]{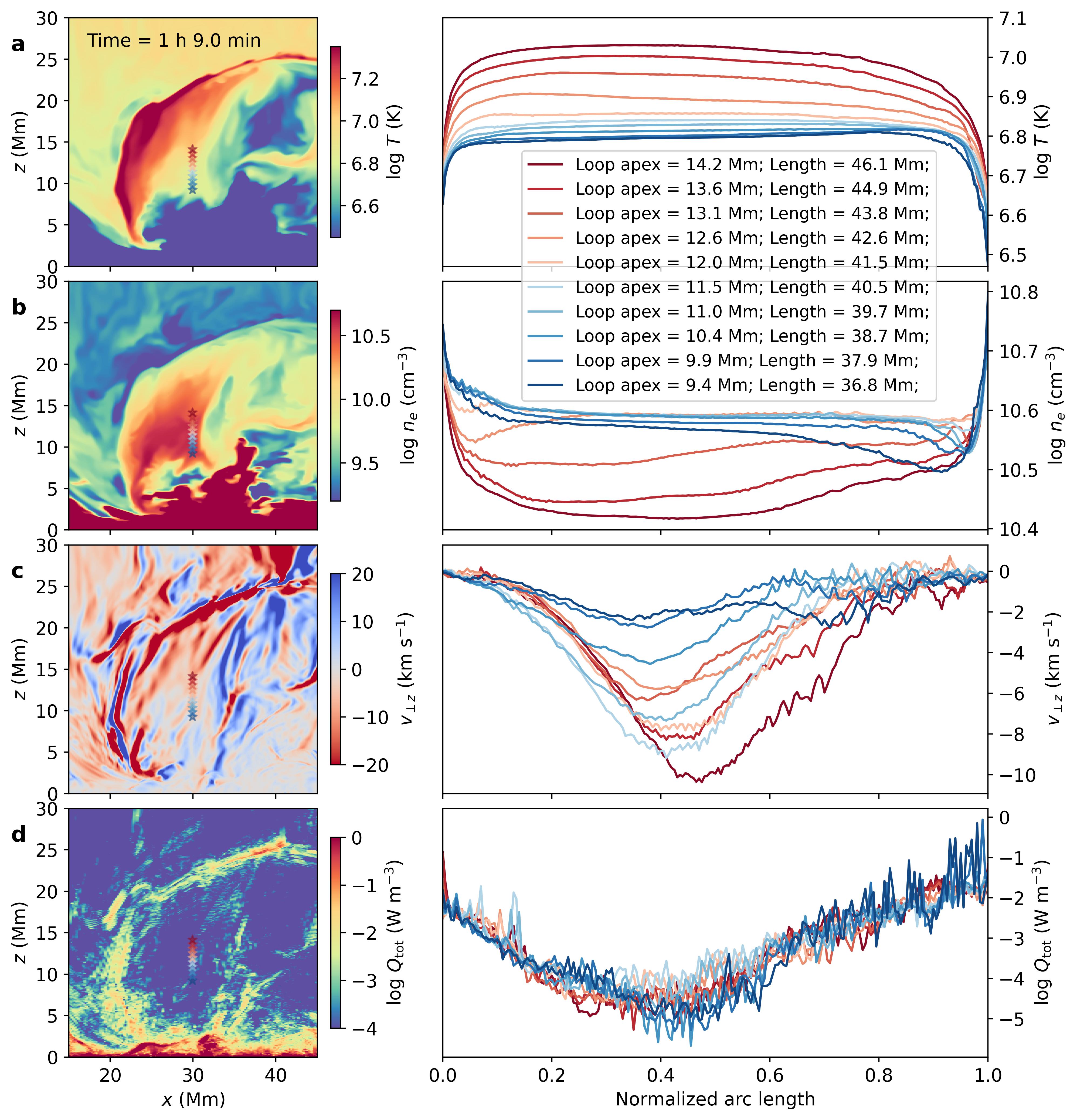}
    \captionsetup{labelformat=extended}
    \caption{\textbf{Quantitative distributions of selected physical parameters.} Distributions of temperature (panel a), number density (panel b), $z$-component of the velocity perpendicular to the magnetic field (panel c), and volumetric heating rate (panel d) are shown. Left panels depict the 2D distributions of the quantities on the cross section of the hot coronal loops at $y = 43\ \mathrm{Mm}$. The colored stars mark positions of the seed points of the magnetic field lines on the slice, where the color notations are the same as those in the right panels. The seed points are set evenly along the $z$-direction in a length of 5 Mm. The field lines passing through these seed points are plotted in Extended Data Fig. 7a. The right panels show distributions of the quantities along the ten field lines.}
    \phantomsection
    \label{efig:flows_lines_chart}
\end{figure}

\begin{figure}[h]
    \centering
    \includegraphics[width=1.0\textwidth]{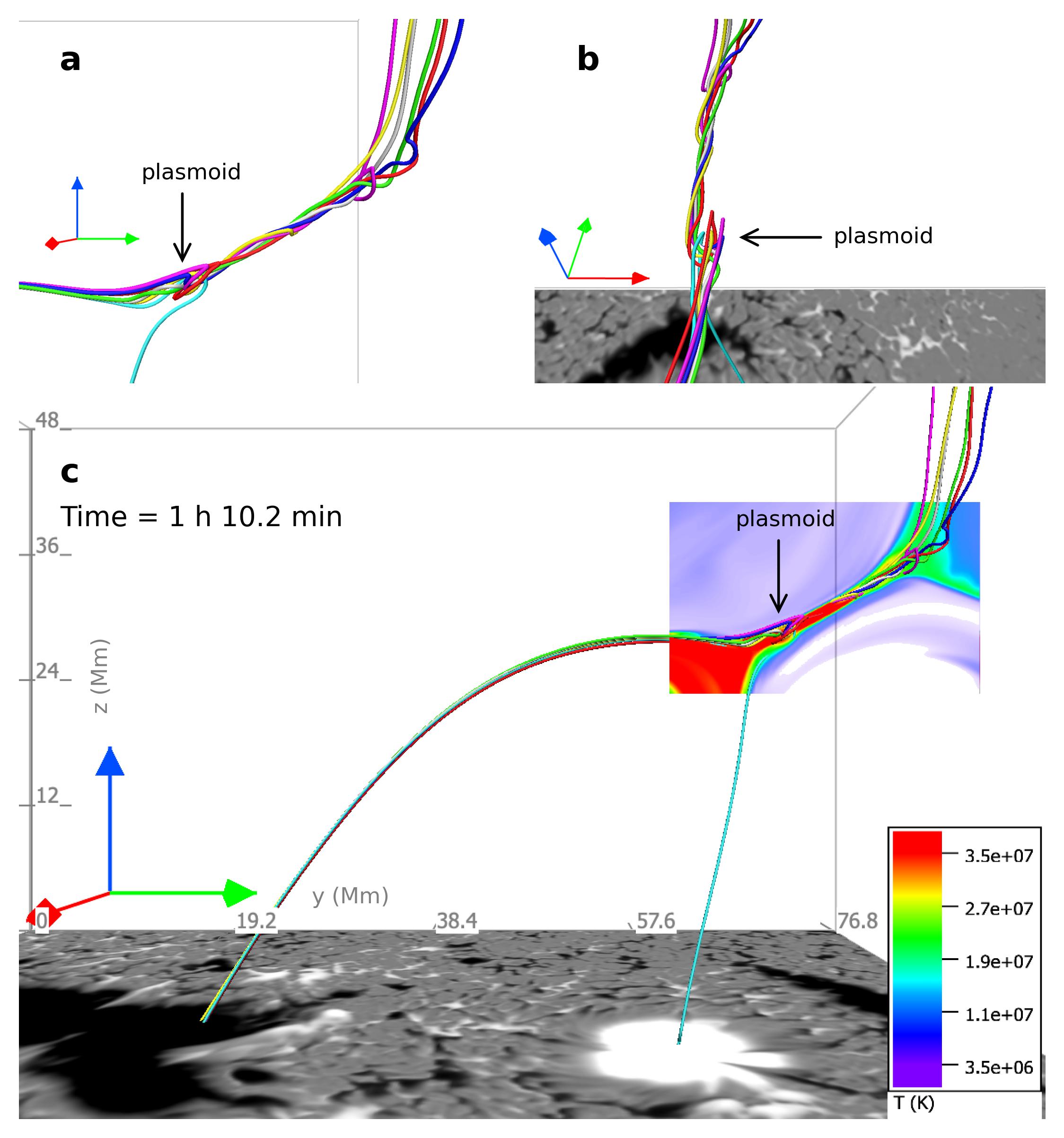}
    \captionsetup{labelformat=extended}
    \caption{\textbf{3D views of the magnetic field lines corresponding to the plasmoid structure.} \textbf{a,} A zoom-in view of the mini flux rope within the current sheet. The viewpoint aligns with that of panel (c), both oriented along the $x$-direction. The red, green and blue arrows represent $x$-, $y$- and $z$-axes, respectively. \textbf{b,} A zoom-in view from another perspective. \textbf{c,} An overview of the mini flux rope within the current sheet. The colored slice represents the 2D distribution of temperature with the same domain as in \mbox{Fig. \ref{fig:fig4}e}.}
    \phantomsection
    \label{efig:vapor_plasmoid}
\end{figure}






\end{document}